\documentclass{iopart}
\usepackage{graphicx}

\begin{document}

\title{Interference of Spontaneous Emission of Light from two Solid-State Atomic Ensembles}
\author{M Afzelius$^{1}$, M U Staudt$^{1}$, H de Riedmatten$^{1}$, C Simon$^{1}$, S R Hastings-Simon$^{1}$, R Ricken$^{2}$, H Suche$^{2}$, W Sohler$^{2}$ and N Gisin$^{1}$}
\address{$^1$ Group of Applied Physics, University of Geneva, 1211 Geneva, Switzerland}
\address{$^2$ Angewandte Physik, University of Paderborn, 33095 Paderborn, Germany}
\ead{mikael.afzelius@physics.unige.ch}

\begin{abstract}
We report an interference experiment of spontaneous emission of
light from two distant solid-state ensembles of atoms that are
coherently excited by a short laser pulse. The ensembles are
Erbium ions doped into two LiNbO$_3$ crystals with channel
waveguides, which are placed in the two arms of a Mach-Zehnder
interferometer. The light that is spontaneously emitted after the
excitation pulse shows first-order interference. By a strong
collective enhancement of the emission, the atoms behave as ideal
two-level quantum systems and no which-path information is left in
the atomic ensembles after emission of a photon. This results in a
high fringe visibility of 95\%, which implies that the observed
spontaneous emission is highly coherent.
\end{abstract}

\maketitle

\section{Introduction}
Spontaneous emission from atoms is one of the most commonly
observed quantum effects in physics
\cite{MANDELandWOLF,Scully1997}. Inherent to the emission is the
randomness of the spontaneous process. Therefore one may think
that the spontaneous emission cannot be phase coherent with
respect to an excitation laser, which is a point of view often
repeated in textbooks on optics and lasers. However, the coherence
properties of spontaneous emission have been thoroughly discussed
theoretically, i.e. in the context of resonance fluorescence
\cite{MOLLOW69a,KIMBLE76a,Cohen-Tannoudji1977}, superradiance
\cite{DICKE54a}, or optical free-induction decay
\cite{MANDELandWOLF,BREWER72a,Jodoin1974}. In the case of
resonance fluorescence experiments, for instance, subnatural
linewidths have been observed using heterodyne measurements
\cite{Westbrook1990,Hoffges1997b}, which demonstrates that
resonance fluorescence emission can be highly coherent.

Another way of exploring the phase coherence of spontaneous
emission is by performing interference experiments. Yet few
reports on interference of spontaneous emission from atoms have
been published. A pioneering interference experiment in this
context was performed by Eichmann et al. \cite{EICH93a}, where two
trapped $^{198}$Hg$^+$ ions played the role of slits in a Young's
double-slit experiment. At low laser intensities, they observed
interference fringes in the resonance fluorescence from the two
ions. This experiment has been thoroughly discussed
\cite{Scully1997,ITANO98a,KIFF06a,AGAR02a,FEAGIN06a} and an
interesting which-path interpretation has been given
\cite{ITANO98a}. There it was argued that if excitation and
emission take place in a closed two-level system, then the
information about which path the photon took is erased from the
atoms (quantum erasure \cite{SCULLY82a}), and as a result
interference is observed. However, if the emission leaves the atom
in a state different than the initial state one could in principle
know by which path the photon passed, and the interference pattern
disappears. These two cases were explored in Ref. \cite{EICH93a}
by detecting either $\pi$- or $\sigma$-polarized light, where
interference was observed in the former case but not in the
latter. However, the visibility when observing $\pi$-polarized
light was limited by a number of factors, including spontaneous
Raman scattering to other states than the initial one.

Here we present an experiment where Erbium ions doped into two
LiNbO$_3$ crystals, i.e. solid-state atomic ensembles, placed in
the two paths of a Mach-Zehnder interferometer, are excited by a
coherent laser pulse. We show that the spontaneous emission
following the pulsed excitation, detected at the output of the
interferometer, exhibits first-order interference with high
visibility. The use of macroscopic atomic ensembles collectively
enhances the spontaneous emission in the forward direction on the
transition connected by the coherent excitation laser
\cite{MANDELandWOLF,Scully2006}. This type of emission is also
known as optical free-induction decay (FID) emission
\cite{MANDELandWOLF,BREWER72a}, which has a $N^{2}$ intensity
dependence on the number of atoms $N$ since all atoms are
initially spontaneously radiating in phase. The collective $N^{2}$
enhancement of the emission probability means that the spontaneous
emission on the excited transition will dominate over emissions on
other transitions. The ensembles can then be considered as being
composed of ideal two-level atoms, as required for observing
high-visibility interference from the which-path argument
mentioned above. Due to the long coherence time of the optical
transition we used, the collective spontaneous emission can be
clearly separated in time from the excitation pulse making it
possible to detect it. The resulting interference fringe
visibilities are excellent (V = 95\%), clearly demonstrating that
spontaneous emission of light can be coherent.

Our experiment relates closely to an experiment proposed by Mandel
\cite{Mandel1983}. There he supposed that two-level atoms in two
independent ensembles were prepared in two coherent superposition
states with relative phase $\Delta\phi$. This could be done by
exciting the ensembles with two coherent laser pulses having a
phase difference $\Delta\phi$, as in the experiment discussed in
this paper. Mandel \cite{Mandel1983} then found that the
spontaneous emission from the two ensembles detected on a screen
would show first-order interference, provided that the phase
difference $\Delta\phi$ remained sufficiently stable. The main
difference as compared to our experiment is that we detect the
emission in a single spatial mode and we instead observe
first-order interference by slowly scanning the phase difference
$\Delta\phi$.

In comparison with the interference experiment of Ref.
\cite{EICH93a}, a main novelty of this experiment is the use of
macroscopic solid-state ensembles having long optical coherence
times and the resulting collective enhancement of the spontaneous
emission. These features allow us to observe much higher fringe
visibilities. We also have a significantly larger spatial distance
between the ensembles ($\sim$7 cm compared to $\sim$5 $\mu$m).
Another important difference is the pulsed excitation in our
experiment, as compared to the continuous excitation in resonance
fluorescence experiment \cite{EICH93a}. This results in a clear
separation in time of the excitation pulse and the detection,
which means that the atoms evolve freely after excitation until
spontaneous emission takes place. This also avoids some additional
complications related to resonance fluorescence experiments, where
frequency side bands appear in the emission at high laser
intensities (the Mollow triplet)
\cite{MOLLOW69a,KIMBLE76a,Cohen-Tannoudji1977}.

\section{The Interference Experiment}
\begin{center}
\begin{figure}
\includegraphics[width = 10 cm]{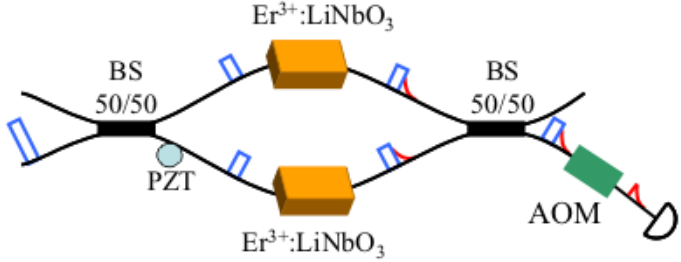}
\caption{Experimental setup for observing interference of
collective spontaneous emission from two solid-state atomic
ensembles. The excitation light pulse is created by intensity
modulation of a cw external-cavity laser diode using a combination
of acousto- and electro-optic modulators (not shown). The laser
pulse is split into two pulses at a fiber 50/50 beam splitter
(BS), which coherently excites the Erbium ions doped into two
LiNbO$_3$ waveguides. These are placed inside a pulse-tube cooler
at a temperature of 3 Kelvin and separated by 7 cm. The collective
spontaneous emission from the Erbium ensembles is then combined at
another 50/50 fiber BS, forming a balanced Mach-Zehnder
interferometer. A piezo-electric transducer (PZT) is used to
control the phase of the interferometer. In front of the detector,
the acousto-optic modulator (AOM) serves as an optical gate to
suppress the excitation pulse, in order to avoid saturating the
detector.} \label{setupFID}
\end{figure}
\end{center}

An excitation pulse created by intensity-modulating the cw-light
from an external-cavity diode laser excited Erbium ions doped into
two LiNbO$_3$ inorganic crystals placed in the arms of an
Mach-Zehnder interferometer, see Fig. \ref{setupFID}. The Erbium
ions absorbing within the frequency bandwidth of the laser pulse
were coherently excited, creating a macroscopic dipole moment in
the two samples. Owing to the long optical coherence time of the
transition (see below), a strong collective spontaneous emission
(or FID emission) was observed after the excitation pulse (see
Fig. \ref{fighigh}). By collective we mean that the spontaneous
emission is enhanced by constructive interference in the forward
direction along the spatial mode of the excitation laser, leading
to an emission probability proportional to $N^2$, where $N$ is the
number of atoms in the excitation volume
\cite{MANDELandWOLF,Scully2006}. In general the FID emission
decays due to inhomogeneous or homogeneneous dephasing processes,
as seen in Fig. \ref{fighigh}. The collective enhancement only
takes place in the forward direction on the excited transition,
where an optical coherence has been induced. The spontaneous
emission into other spatial modes and on other transitions is
non-collective, therefore leading to an emission probability only
proportional to the number of atoms \emph{N}. We emphasize that,
while non-collective spontaneous emission on the excited
transition can be coherent, emission on other transitions is
entirely incoherent.

The Erbium ions were excited on the near-infrared transition
$^4$I$_{15/2}$-$^4$I$_{13/2}$ at 1532 nm \cite{Gruber2004}. In
general, rare-earth-metal-ion-doped solid-state materials have
spectrally narrow absorption lines and excellent optical coherence
properties at low temperatures ($<$4K)\cite{Sun2005}. The Erbium
ions can then be considered as a frozen gas naturally trapped in
the crystalline host. In Er$^{3+}$:LiNbO$_3$ the
$^4$I$_{15/2}$-$^4$I$_{13/2}$ absorption spectrum is
inhomogeneously broadened to about 250 GHz by site-to-site
variations in the static interaction between Er$^{3+}$ ions and
the LiNbO$_3$ host \cite{SUN02a}. The homogeneous linewidth,
however, is of the order of 30 kHz at the experimental temperature
of $\sim$3 K \cite{STAUDT07a}, which corresponds to an optical
coherence time of $T_2$ $\sim$10 $\mu$s. To obtain this coherence
time, a small magnetic field ($>$0.1 Tesla) must be applied along
the crystal c-axis to reduce magnetic spin interactions in the
material, which otherwise lead to fast optical decoherence
\cite{Sun2005,SUN02a}.
In our experiment we used two Er$^{3+}$-doped LiNbO$_3$ waveguides
(20 mm long and 10 mm wide). The LiNbO$_3$ crystal surfaces were
doped with Erbium ions by indiffusion, and optical channel
waveguides (Ti-indiffused) were integrated on the surface
\cite{STAUDT07a,Baumann1996}, allowing single-mode waveguiding of
the 1.5 $\mu$m light through the entire interferometer. The
waveguides were not identical because waveguide II had two times
higher Erbium doping concentration than waveguide I (waveguide I:
$4\cdot10^{19}$/cm$^3$ surface concentration before indiffusion),
resulting in a higher absorption in waveguide II. The fiber in one
of the arms of the interferometer was partly coiled around a piezo
element, which allowed control of the phase difference of the
interferometer. The entire interferometer was installed in a
pulse-tube-refrigerator. The Er$^{3+}$:LiNbO$_3$ waveguides were
placed on the low-temperature level for cooling to about 3 K,
whereas the 50/50 fiber beam splitters were placed at ambient
temperature to ensure proper functioning. As a result, the arms of
the interferometer were 2.63 m long. Note that there was then a
temperature gradient of about 300 K across the interferometer.
Since the fibers in the interferometer were not polarization
maintaining, it was necessary to project the axis of polarization
of the emission from the two ensembles onto a common axis. This
was done by placing a fiber polarization controller (FPC) and a
fiber polarizer (FP) in front of the detector (all outside the
pulse-tube cooler). The total loss in each arm of the
interferometer was roughly 14 dB, mostly due to input and output
couplings of light between the single-mode fibers and waveguides.
The AOM serving as optical gate and the FPC+FP introduced another
8 dB loss between the output of the interferometer and the
detector.

To characterize the maximum visibility of the interferometer, we
performed an interference experiment using a cw laser tuned off
the Erbium resonance (to 1550 nm). With the cooling system turned
on, we obtained a maximum visibility of about 92$\%$, whereas with
the cooling system turned off about 100$\%$ was obtained (the
absolute error estimated from several measurements were about
1$\%$ in both cases). In the former case, the visibility was
clearly limited by phase noise introduced by vibrations in the
pulse-tube cooler. Note that the experiment was performed at a
repetition rate of 13 Hz, which was found to limit the effect of
vibrations on the phase noise. The long-term passive stability of
the interferometer was then good enough to perform interferometric
measurements over tens of minutes.

The experiment was carried out both in a high and low excitation
regime. In the former, a strong excitation pulse was used, such
that a classical detector could be used to detect the emission at
the output of the interferometer. This resulted in a good
signal-to-noise ratio and shorter integration times for each point
on the interference curve. In the latter, we reduced the
excitation pulse energy such that a single-photon detector could
be used for detection. Although longer integration times were
needed to obtain good signal-to-noise ratios, this experiment more
clearly emphasizes the quantum nature of the spontaneous emission.
In both cases, however, the experiment can be explained in terms
of coherent states of light (bright or weak). In order not to
saturate the detector, we used an acousto-optic modulator as an
optical gate before the detector. In the high excitation
experiment, the optical gate of 1 $\mu$s was opened 130 ns after
the excitation pulse. In the low excitation experiment, the
optical gate was opened 700 ns after the excitation pulse, whereas
the 100 ns detection window of the single-photon detector was
opened 1 $\mu s$ after the excitation pulse.

\section{Results and Discussion}
\begin{center}
\begin{figure}
\includegraphics[width = 10 cm]{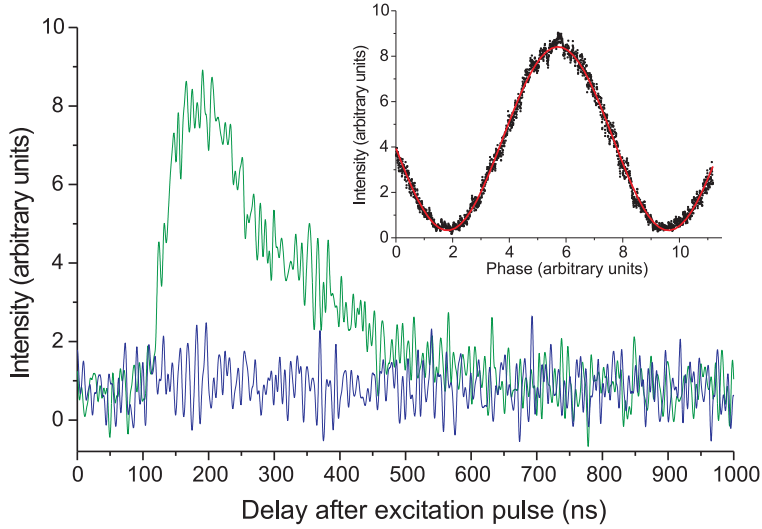}
\caption{Interference of collective spontaneous emission in the
high excitation regime. The graph shows the constructive (green)
and destructive (blue) interference signals as a function of time
after the end of the 2 $\mu$s long excitation pulse. The signals
were detected by a classical detector. The optical detection gate
was opened 130 ns after the excitation pulse, such that the signal
the first 130 ns represents the detector noise level. Inset: The
area under the signal (detector noise subtracted) as a function of
phase difference. In this case, the measured interference
visibility is 93$\pm1.5\%$.} \label{fighigh}
\end{figure}
\end{center}
\begin{center}
\begin{figure}
\includegraphics[width = 10 cm]{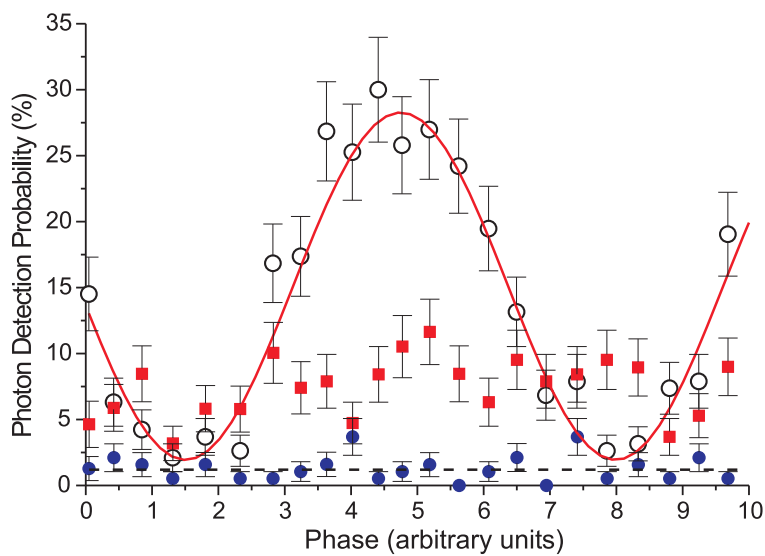}
\caption{Interference in the low excitation regime. The points
represent the measured detection probability as a function of the
phase of the interferometer for three different experimental
situations. For black open circles, the excitation laser is on
resonance (1532 nm) and the light emitted by both ensembles is
detected and shows clear interference. The solid line is a
sinusoidal fit, leading to a net visibility of 95$\pm5\%$. For
blue filled circles, light from both ensembles is detected, but
the excitation laser is far out of resonance (1550 nm). In this
case the detection probability drops to the level of dark noise of
the single-photon detector (represented by the dashed line) . For
red filled squares, the laser is on resonance, but only the light
emitted by one ensemble is detected, and as expected no
interference fringes are observed.} \label{figlow}
\end{figure}
\end{center}

In the high excitation regime, the optical pulse had a duration of
2 $\mu$s and a peak power of 2 $m$W ($4\cdot10^{10} $ photons per
pulse) at the entrance of the interferometer (see above concerning
losses in the interferometer).  In Fig. \ref{fighigh} the
collective spontaneous emission is shown with the phase difference
of the interferometer tuned to constructive and destructive
interference. The maximum spontaneous emission signal (in front of
the detector) was about 90 nW at constructive interference,
corresponding to $7\cdot10^4$ photons/100ns. The decay of the
signal in this case is approximately $\sim$150 ns, which
corresponds rather well to that obtained by numerically solving
the Maxwell-Bloch equations using parameters corresponding to the
current experiment. By measuring the area of the signal as a
function of phase difference we obtained clear interference
fringes, as shown in the inset of Fig. \ref{fighigh}.
\\In the low
excitation regime, the peak power of the pulse was reduced to 10
$\mu$W ($2\cdot10^8$ photons per pulse). Since free-induction
decay emission is a third-order non-linear process
\cite{BREWER72a}, this reduction was sufficient to make it
possible to detect the emission using a single-photon detector. By
recording the detection probability while scanning the phase
difference of the interferometer, we observed interference fringes
of $95\pm 5\%$ visibilities (detector noise subtracted), as shown
in Fig. \ref{figlow}. Note that this result is within the
technical limit of 92$\%$ set by phase noise in the interferometer
(see previous section). The detection probability at constructive
interference was 30$\%$, which translates to 3 photons per 100 ns
detection window in front of the detector (taking into account the
10$\%$ detection efficiency). The detector noise level was 1.2$\%$
due to dark counts. We verified that the photons detected were
indeed emitted from the ensembles, and not laser light leaking
through the intensity modulators. This was done by tuning the
laser wavelength outside the optical resonance (to 1550 nm), such
that no atoms were excited. As expected, the detection probability
then dropped to the noise level of the detector, see Fig.
\ref{figlow}, which proves that the intensity modulators provided
good enough extinction to observe the few-photon spontaneous
emission. To show that the interference is due to emission from
both ensembles in the two arms, we "turned off" the collective
emission from one of the arms by removing the magnetic field on
the corresponding sample. This reduces the optical coherence time
by several orders of magnitude, which in turn drastically shortens
the decay of the collective signal. The emission from this arm was
then at a non-detectable level at the time of the single-photon
detection window. As expected, the photon detection probability
then showed no interference as a function of phase difference (see
Fig. \ref{figlow}), and it dropped to about one fourth of the
constructive interference signal observed with the collective
emission "turned on" in both arms.

In order to understand this experiment, one may follow a single
photon going through the interferometer. After the first beam
splitter, the photon is in a state of superposition of being in
the two arms. The photon is then absorbed by the two ensembles,
which are ideally ensembles of two-level quantum systems in
resonance with the photon. The photon is now stored in both
ensembles as a delocalized single excitation. After some time the
photon is spontaneously emitted, the two modes are combined on the
second beam splitter, and the photon is thereafter detected by the
single-photon detector. Only if the emitted photon is phase
coherent with the absorbed photon can one observe perfect
interference visibility. Hence the experiment presented here
clearly and directly demonstrates the coherent nature of the
observed spontaneous emission.
As discussed above, an important condition for observing
interference is that only emission on the excited transition is
observed. The detection of a photon emitted on another transition
implies that the atom is left in another state than the initial
one, and which-path information is left in the Erbium-ion
ensembles. In the case of Er$^{3+}$:LiNbO$_3$, the energy
structure is very rich due to the different crystal-field, Zeeman
and hyperfine levels \cite{Gruber2004}. In this experiment, the
atoms were excited from the lowest crystal-field (CF) level in the
electronic ground state $^4$I$_{15/2}$(0) to the lowest CF level
in the first electronically excited state $^4$I$_{13/2}$(0). It is
the collective enhancement on the excited
$^4$I$_{13/2}$(0)-$^4$I$_{15/2}$(0) transition that allows us to
discriminate against emission to other states (particularly to
other CF levels in the ground state). In this way, the atoms act
as ideal two-level quantum systems, and no information about the
previous excitation is left within the atoms. Note that no
spectral filtering was used in the experiment.

In the introduction we mentioned that a theoretical calculation of
a thought experiment closely related to this experiment has been
published by Mandel \cite{Mandel1983}. In particular, he
calculates the expected visibility as a function of the number of
atoms in each ensemble and the degree of excitation of the atoms.
In the case when the number of atoms in each ensemble is the same,
$N$, the theoretical visibility is \cite{Mandel1983}

\begin{center}
$V=\frac{N\cos^2\frac{1}{2}\theta}{1+(N-1)cos^2\frac{1}{2}\theta}$
\end{center}

where $\theta$ is the normal pulse area. If $N\gg1$, as in the
experiment presented here, the visibility is close to 1, almost
independently of the excitation $\theta$, except when all atoms
are excited ($\theta=\pi$). If $N=1$, however, as in the case of
the two trapped ions in the experiment presented in Ref.
\cite{EICH93a}, the visibility becomes strongly dependent on the
degree of excitation $\theta$, and only at low excitation
$\theta\approx0$ does one observe perfect visibility
\cite{Scully1997,Mandel1983}. The use of large atomic ensembles
presents an advantage also from this point of view.

Interference of light emitted by atoms has also been studied from
a more applied perspective, because it plays a central role in
quantum information research. In quantum networks, for instance,
quantum states of light stored and retrieved from independent
atomic memories would need to interfere with very high fringe
visibilities \cite{Duan2001}. In this context conditional
first-order quantum interference of Raman photons produced by
four-wave mixing in two three-level ensembles of cold atoms has
been reported \cite{Matsukevich2004,Chou2005}. There the emission
of the interfering photons is also collectively enhanced, but
simultaneous with the excitation laser (which is at a different
frequency). The observation of interference is conditional on the
detection of a first photon which projects the ensembles in a
state with a delocalized collective atomic spin excitation. Note
also that the fundamental effect of collective spontaneous
emission observed in this paper is at the heart of photon echo
techniques \cite{MANDELandWOLF}, which are being studied in the
context of photonic quantum storage
\cite{STAUDT07a,Moiseev2001,Alexander2006,Staudt2007}.
\\
\section{Conclusions}
To conclude, we have demonstrated high-visibility interference of
the spontaneous emission of light from two spatially separated
solid-state atomic ensembles. The high contrast observed has been
made possible by the strong collective enhancement of the
spontaneous emission which causes the multi-level Erbium ions to
behave as ensembles of ideal two-level systems. This clearly
demonstrates that light spontaneously emitted from separated
atomic systems can be highly coherent, provided that the initial
excitation is coherent and that no which-path information is left
in the atoms.

\section{Acknowledgements}
This work was supported by the Swiss NCCR Quantum Photonics and by
the European Commission under the Integrated Project Qubit
Applications (QAP). The authors thank Nicolas Sangouard and
Wolfgang Tittel for stimulating discussions.

\section{References}
\bibliography{references}
\bibliographystyle{unsrt}

\end{document}